\documentclass[12pt]{article}

\usepackage{amsmath,amssymb,epsfig}

\def\be{\begin{equation}}
\def\ee{\end{equation}}

\def\Ep{E_+}

\begin{document}

\begin{center}
{\Large\bf Properties of kinematic singularities}
\vspace{.3in} \\ 
{\bf A A Coley}, 
\\Department of Mathematics \& Statistics, Dalhousie University,\\
Halifax, Nova Scotia, Canada B3H 3J5
\\Email: aac@mathstat.dal.ca
\vspace{.1in}
\\ {\bf S Hervik}, 
\\Dept. of Mathematics and Natural Sciences, 
University of Stavanger, N-4036 Stavanger, Norway
\\Email: sigbjorn.hervik@uis.no
\vspace{.1in}
\\ {\bf W C Lim},
\\Albert-Einstein-Institut, Am M{\"u}hlenberg 1, D-14476 Potsdam, Germany
\\Email: wclim@aei.mpg.de
\vspace{.1in}
\\{\bf M A H MacCallum}
\\School of Mathematical Sciences, Queen Mary University of London,\\
E1 4NS, UK
\\Email: m.a.h.maccallum@qmul.ac.uk
\vspace{0.2in}

\today
\end{center}

\begin{abstract}

The locally rotationally symmetric tilted perfect fluid Bianchi type V
cosmological model provides examples of future geodesically complete
spacetimes that admit a `kinematic singularity' at which the fluid
congruence is inextendible but all frame components of the Weyl and
Ricci tensors remain bounded.  We show that for any positive integer $n$ there are
examples of Bianchi type V spacetimes admitting a kinematic
singularity such that the covariant derivatives of the Weyl and Ricci
tensors up to the $n$-th order also stay bounded. We briefly discuss
singularities in classical spacetimes.

\end{abstract}

\

\newpage

\section{Introduction}

A cosmological model ($\mathcal{M},\mathbf{g},\mathbf{u}$) is defined
by specifying the spacetime geometry, determined by a Lorentzian
metric $\mathbf{g}$ defined on the manifold $\mathcal{M}$, a
family of fundamental observers, whose congruence of worldlines is
represented by the 4-velocity field $\mathbf{u}$, and some appropriate
matter content. The covariant
derivative $u_{a;b}$ of the 4-velocity field can be decomposed into
\emph{kinematic} variables according to \be u_{a;b} = \sigma_{ab} +
\omega_{ab} + H (g_{ab} + u_a u_b) - \dot{u}_a u_b, \ee where
$\sigma_{ab}$ is the rate of shear tensor, $\omega_{ab}$ is the
vorticity tensor, $H$ is the Hubble scalar, and $\dot{u}_a$ is the
acceleration vector.

In a recent paper \cite{CHL} the future\footnote{`Future' meaning
`in the direction of expansion'.} asymptotic dynamics as experienced
by the perfect fluid observer in spatially homogeneous (SH) Bianchi
cosmologies with a tilted perfect fluid with linear equation of state
$p=(\gamma-1)\mu$ was examined. It was found that given a Bianchi type
and a large enough value of the equation of state parameter, $\gamma$,
the perfect fluid observer can encounter a singularity in the sense
that the fluid congruence becomes inextendible after a finite proper
time. This is not a true spacetime singularity, since the causal
geodesics are complete. Indeed, observers moving orthogonally to the
SH hypersurfaces, and therefore geodesically, would not encounter a
spacetime singularity in the future \cite{Rendall}. The fluid does not
follow geodesics: it is accelerated.  The pathology of the flow
congruence is therefore called a kinematic singularity\footnote{So
called at some risk of confusion with true spacetime singularities
as defined in section~\ref{Discussion} below:
the word `singularity' is retained here because the two concepts have
in common the idea of incomplete inextendible causal curves, while
differing in whether these must be geodesic or can accelerate.}
\cite{LCH}: it is accompanied by an extreme tilt limit (i.e.\ the
fluid observers' motion becomes asymptotically null with respect to
the SH observers). It has the properties that (i) the proper time
needed to reach the kinematic singularity is finite, (ii) the Hubble
scalar $H$ and some other kinematic quantities diverge, and (iii) the
matter density tends to zero (i.e.\ the kinematic singularity is not a
matter singularity). Since the source is a perfect fluid with a linear
equation of state, this last means that all components of the Ricci
tensor in the comoving frame tend to zero.

The {\em Weyl parameter}, $\mathcal{W}$, is defined by
\cite{WHU99} \be \label{Weyl_param}
    \mathcal{W} = \frac{W}{H^2},\quad
    \text{where}
    \quad
    W^2 = \frac16(E_{ab}E^{ab} + H_{ab}H^{ab}).
\ee
The phenomenon of {\em Weyl curvature dominance} occurs when
$\mathcal{W} \rightarrow \infty$, which can happen in Bianchi type
VII$_0$ and VIII cosmologies \cite{WHU99,VII}.
The Weyl parameter can be defined with respect to
different 
invariantly defined systems of observers and,
particularly in Bianchi models with a tilted perfect fluid,
with respect to
the fluid observers and the spatially homogeneous observers.
Although the limits of the Weyl parameter
along these two congruences of worldlines are in principle different, it
has been conjectured that in general they both tend to zero,
constant, or infinity together \cite{LCH}. However, note that this
does not occur in the exceptional locally rotationally-symmetric 
(LRS) Bianchi type V models, due to cancellation of terms.

Suppose that the worldlines of a congruence are incomplete and
inextendible. For brevity, we shall talk about limiting behaviour on
the curves as if it were possible to attach endpoints to them,
although there are difficulties in doing so in such a way as to
construct a sensible boundary to the spacetime \cite{TCE}.  If a
curvature scalar blows up as the endpoint is approached, the endpoint
may be a true spacetime singularity, as discussed in section
\ref{Discussion} (the alternative is that the blow up occurs at
infinity along causal geodesics). If the matter (or Ricci) tensor
blows up, we call this a matter singularity, while if only the Weyl
tensor is singular we call it a Weyl singularity
\footnote{This definition of a Weyl singularity
is a more natural terminology
for the case of a true spacetime singularity
-- it differs from the definition of a
`Weyl singularity' utilized
in \cite{LCH}, which we shall refer to here as
Weyl blow-up in finite proper time.}  
 (we could use the
term `conformal singularity', following \cite{CE79}, but we wish to
avoid confusion with the `conformal singularity' in the context of
isotropic singularities \cite{AT99}).

If at least one component of the Weyl curvature tensor (with respect
to the orthonormal frame of an observer travelling along the
congruence) diverges, we call this {\em Weyl blow-up}{\footnote{Note
that it is possible that all Weyl scalar invariants converge, and the
Weyl tensor is non-singular, but that Weyl blow up occurs, which
implies that one observer may experience Weyl blow-up while another
does not. This can, and in general will, happen when the Weyl tensor
invariants have a non-zero limit along an observer's worldline and in
the limit that observer's motion becomes lightlike with respect to the
frame fixed by the Weyl tensor (e.g., see Chapter 4 of
\cite{SKMHH}).}}. Another possibility is that the matter tensor and
Weyl tensor both converge as the endpoint is approached, but some
kinematic quantities (most importantly $H$) diverge, i.e.\ a kinematic
singularity occurs. Thus, relative to a timelike congruence, a
kinematic singularity is characterized by the blow-up of one or more
kinematic variables in finite proper time, while all components of the
Weyl and Ricci tensors remain bounded.

On approach to an extreme-tilt limit, the fluid observers in the
tilted Bianchi cosmologies may or may not experience Weyl blow-up, and
so the limit may or may not be a kinematic singularity \cite{LCH}.
The LRS Bianchi V models provide examples of
non-spacetime singular models both with Weyl blow-up and with a
kinematic singularity.  When studying the properties of a true
singularity attention is usually focussed on the behaviour of the Weyl
tensor and the Ricci tensor. However, there are examples in which Weyl
blow-up does not indicate a singularity and kinematic quantities are
useful in indicating a kinematic singularity. An important related
question, which we shall address here, is whether a kinematic
singularity is always associated with the divergence of
the components of the covariant derivatives of the Riemann tensor.

\section{LRS Bianchi type V model}

The LRS Bianchi type V model provides examples of non-spacetime
singular models which are geodesically complete and show
Weyl blow-up and kinematic
singularities  (and are such that all scalar invariants
involving the Weyl and Ricci tensors, including mixed invariants,
are bounded). 

First, we give a brief sketch of the derivation of the
LRS Bianchi V limits near the future asymptotic state $M^-$.
We use the dimensionless time variable $\tau$, defined by
equation (2.10) of \cite{HewWai92}:
\[
	\frac{dt}{d\tau} = \frac{1}{H},
\]
where $t$ is the proper time along the spatially homogeneous congruence.
Using the evolution and constraint equations (2.11) and (2.12) of
\cite{HewWai92}, we determine the linearized equations at $M^-$
for the following variables:
\[
        H, \Sigma_+, \Omega, 1-v^2, 1-A^2.
\]
We then compute the decay rates for $E_+$ and $\mathcal{W}$ through
their algebraic expressions
and use the boost formulae to obtain the decay rates for $\hat{H}$ and
$\hat{E}_+$.
Finally, we obtain the decay rates for $\hat{W}$ and $\hat{\mathcal{W}}$.

We take this opportunity to correct some computational errors made in \cite{LCH}.	
Equations (21), (23), (27) and (28) should read, respectively,
\begin{align*}
        H &\approx H_0 e^{-\tau},
        \qquad W \approx 2 H_0^2 |\Sigma_{+0}| e^{-6\tau},
        \qquad \mathcal{W} \approx 2 |\Sigma_{+0}| e^{-4\tau}
\\
        E_+ &= (H+\sigma_+) \sigma_+ + \frac16 \frac{\gamma \mu}{G_+} V^2
\\
        \hat{W} &= W \approx 2  H_0^2 |\Sigma_{+0}| e^{-6\tau}
\\
        \hat{\mathcal{W}} &\approx \frac98(2-\gamma)^2 \Gamma_0^{-2}
         |\Sigma_{+0}| \exp\left( \frac{-2(3\gamma-2)}{2-\gamma}\tau\right).
\end{align*}
The computational errors do not affect the conclusions of \cite{LCH}.

In \cite{LCH} it was shown that in the exceptional
LRS Bianchi V models $H \rightarrow 0, W \rightarrow 0,
\mathcal{W} \rightarrow 0$ as $\tau \rightarrow \infty$ for $6/5 <
\gamma <2$. However, in the fluid frame $\hat{H}\rightarrow
\infty$ as $\tau \rightarrow \infty$ for $4/3 < \gamma <2$, but
$\hat{W} \rightarrow 0, \hat{\mathcal{W}} \rightarrow 0$ for $6/5
< \gamma <2$ and hence do not blow up. Therefore, in this model
there are  kinematic singularities.

In \cite{CE79} it was claimed that the future
singularity in the LRS Bianchi type V model is a  `conformal
singularity'
(figure 7, right side of 7(i) and 7(ii), the bottom spacetime diagram and table).
As mentioned in \cite{LCH},
this is not the case, since the Weyl components tend to zero.
What happens instead is that some kinematic variables blow up in finite time.%
\footnote {The stability of the properties of
the Bianchi V models (within the class of all spatially homogeneous
models) has been studied by Siklos \cite{Siklos}.} Singularities were
classified in \cite{EK74} as `curvature singularities', `intermediate
singularities' and `locally extendible singularities'.  Weyl blow-ups
where at least one of the Weyl scalars also blows up correspond to
curvature singularities, Weyl blow-ups with bounded Weyl scalars (and
Ricci scalars) correspond to intermediate singularities, and kinematic
singularities correspond to locally extendible singularities, or
points at infinity. In our examples, (i) the fluid worldlines are
certainly {\em not} extendible, due to the blow up of the kinematic
variables, and are thus not `locally extendible' in the sense of
Clarke \cite{C73}, (ii) we wish to use terminology that better
describes the type of behaviour under discussion.

The kinematic quantities are given by ratios of Cartan invariants,
using a co-moving frame as the canonical one \cite{parmac}: more
specifically, they are given by dividing the derivatives of the
tracefree Ricci tensor by $(\mu +p)$, a multiple of its one
independent non-zero component. Hence, assuming we do not have $(\mu
+p) \equiv 0$, kinematic singularities arise because (i) both $(\mu
+p)$ and the first derivatives tend to zero, but the ratio tends to
infinity, or (ii) $(\mu +p) \rightarrow 0$ while some first derivative
of the matter tensor remains finite and non-zero, or (iii) the first
derivative blows up, possibly while $(\mu +p)$ remains finite. A
kinematic singularity might also be accompanied by other divergences;
e.g.\ a divergence of a component of the first covariant derivative of
the Weyl tensor.
In order to investigate this here we only need to consider the LRS
Bianchi type V model, since more than one possible behaviour occurs in this
particular example.

\subsection{Weyl derivatives}

For higher derivatives of the Weyl scalar, we first note that $H$ has
the slowest decay rate $e^{-\tau}$
among the kinematic variables. Next, we note that a partial derivative with
respect to proper time,
denoted by an overdot, is $H$ times the partial derivative with
respect to $\tau$.
Since the decay rates are exponential in $\tau$, the time derivative
of a variable has the same decay rate as $H$ times
the variable to be differentiated.

Dimensional consistency implies that
the algebraic terms in a covariant derivative must be a single power
of one of the kinematic variables
times the scalar variable (or tensor components) to be differentiated.
Since $H$ is the slowest decaying kinematic variable, the slowest
decay rate for the algebraic term is
the same as the decay rate for the time derivative above.
We can then apply the same reasoning for higher derivatives.

To convert to the fluid frame components, an orthonormal frame
component (not coordinate component) of a rank $n$ tensor obtained by
contraction with a unit vector in the fluid flow plane, will be
multiplied with $\Gamma^n$ due to the boost formula for the frame vectors.

The LRS Bianchi type V model has only one non-zero Weyl component,
$\Ep$, so consequently $\Ep$ is a scalar invariant and its covariant
derivatives are simply derivatives of $\Ep$. 
$\Ep$ has the decay rate $e^{-6\tau}$.
In this model all of
the scalar curvature invariants are bounded: $C^2\equiv
C_{abcd}C^{abcd} \sim \Ep^2$ and ${C_{,1}}^2\equiv
C_{abcd;e}C^{abcd;e} \sim {\dot{\Ep}}^2 \sim ({H\Ep})^2\rightarrow
0$, and ${C_{,n}}^2\equiv
C_{abcd;e{_1}...e{_n}}C^{abcd;e{_1}...e{_n}} \sim ({H^n\Ep})^2
\sim \exp[-2(n+6)\tau] \rightarrow 0$  (the higher derivatives
converge even faster).

Let us consider individual components of the Weyl tensor and its
derivatives in the fluid frame\footnote {Note that Hubble-normalized
quantities such as for example, $\hat{E}_+/\hat{H}^2$, all tend to
zero.}.  
The boost factor $\Gamma$ has the growth rate 
$\exp\left[\frac{(5\gamma-6)}{2-\gamma}\tau\right]$,
and the Hubble scalar $\hat{H}$ of the fluid frame has the growth rate
$\hat{H} \sim \Gamma H \sim \exp\left[\frac{2(3\gamma-4)}{2-\gamma}\tau\right]$.
These growth rates will be key to the blow up of the derivatives.
The first order derivatives of $\Ep$ in the fluid frame have
leading order of $\dot{\hat{C}} \sim \Gamma{H}\Ep \sim \hat{H}\Ep \sim
\exp\left[\frac{4(3\gamma -5)}{2-\gamma}\tau\right]$.  They blow up
if $\gamma > 5/3$.  The $n$-th order derivatives (both covariant and
partial) of $\Ep$ have leading order of $\hat{H}^n \Ep$. They blow up
if
\[
    \gamma>\frac{2(2n+3)}{3(n+1)}.
\]
A sample of $\gamma$ thresholds for each $n$ are given by:
\begin{center}
\begin{tabular}{|c|ccccc|}
\hline
$n$ &  1 & 2 & 3 & 4 & 5 \\
\hline
$\gamma$ &  5/3 & 14/9 & 3/2 & 22/15 & 13/9\\
\hline
\end{tabular}
\end{center}
Another way to say this is that for a fixed $\gamma$, the derivatives
blow up for
\[
    n > \frac{3(2-\gamma)}{3\gamma-4}.
\]

In the Bianchi V example, for any $n$ there exists a range of values
for $\gamma$ such that the Weyl tensor and its first $n$ covariant
derivatives all stay bounded, and other values for which the Weyl
tensor and/or its derivatives at some orders less than $n$ blow
up. Therefore, at a kinematic singularity both behaviours are
possible, and the notion of blow up of kinematical quantities is
useful.

\subsection{Ricci derivatives}

Similarly, for the Ricci scalar, or equivalently, $\hat{\mu}$,
which has the decay rate
$\hat{\mu} \sim \Gamma^{-2} \mu \sim \exp\left[\left(-2\frac{(5\gamma-6)}{2-\gamma}-6\right)\tau\right]$,
the $n$-th order derivatives (both covariant and partial) of $\hat{\mu}$
have leading order of $\hat{H}^n \hat{\mu}$. They blow up if
\[
        \gamma>\frac{4n}{3n-2},
\]
for $n = 3,4,\cdots$. 
We omit $n=1$ and $n=2$ because the $\gamma$ thresholds are $\geq 2$.
A sample of $\gamma$ thresholds for each $n$ is given by:
\begin{center}
\begin{tabular}{|c|ccc|}
\hline
$n$ & 3 & 4 & 5 \\
\hline
$\gamma$ & 12/7 & 8/5 & 20/13\\
\hline
\end{tabular}
\end{center}
Equivalently, the derivatives blow up for
\[
    n > \frac{2\gamma}{3\gamma-4}.
\]

As an example, consider $\gamma = 5/3$. In this case the
components of both the Weyl tensor and the Ricci tensor tend to
zero. The components of the first derivative of the Weyl tensor
are bounded, but the components of the second derivative of the
Weyl tensor are unbounded, while the derivatives of the Ricci
tensor up to third order tend to zero and the components of the
fourth derivative of the Ricci tensor blow up.

\section{Discussion}\label{Discussion}

Due to  these
results, and given the recent tendency to use imprecise
terminology, particularly in the physics literature, perhaps it would be useful
to briefly review singularities in classical general relativity.

The existence of a spacetime singularity is defined by the inextendibility of
causal (that is, null or timelike) geodesics \cite{Geroch,Wald}.
More precisely, a manifold is said to have a spacetime singularity
if its maximal development is geodesically incomplete.
The problem with defining a boundary point of spacetime at which the
singularity can be said to occur is that the
geometry is singular there, so that the singularity cannot be a
regular spacetime point. 
So these points are never in the manifold initially considered, but
the definition requires its maximal development.
This implies that the 
identification of a real
singularity depends on the criteria chosen for extendibility; for example,
if discontinuous extensions were allowed, then the
Friedmann solutions would be extendible and so would not have real
singularities \cite{TCE}.

For the existence of solutions of the geodesic equation, the metric
must be $C^1$ and for these solutions to depend continuously on
initial position and direction, it must be $C^{2-}$ (see
\cite{HawEll73})\footnote{In this and subsequent statements we use
function spaces defined by differentiability and Lipshitz
conditions. Somewhat weaker conditions using Sobolev spaces actually
suffice; e.g., see \cite{TCE}.}.  There are various other
differentiability conditions (on the metric or on the Riemann tensor
components) that could be used (together with other additional
conditions). Perhaps the most widely used inextendibility assumptions
are that the extended metric be $C^N$, where $N$ is the minimum
differentiability needed for the existence of a unique maximal
development of the Einstein equations from a set of initial data
\cite{TCE}. Using this definition the class of non-singular spacetimes
includes at least some shock waves where there is a discontinuity in
the Riemann tensor, but not impulse waves. Indeed, any mathematical
model of inhomogeneous matter model is likely to allow shock waves to
form (and non-analyticity will cause high-enough derivatives of
Riemann tensor to blow up).

Once a spacetime singular point has been constructed, its
properties are of interest. There
has been much work on the occurrence and possible natures of
singularities thus defined \cite{TCE}.
Singularities can be classified according to the boundedness or
divergence of scalar polynomials of the Riemann tensor (and its
covariant derivatives). A scalar polynomial curvature singularity
is the end point of at least one curve on which a scalar
polynomial in the metric and the Riemann tensor takes unboundedly
large values (and hence some Riemann tensor component is
unbounded in {\it any} tetrad, and not just in parallelly propagated
ones). Archetypical examples are the Schwarzschild solution with a
timelike curvature singularity in which the Kretschmann scalar
(which is equivalent to the square of the Weyl curvature in this
case) diverges, and the Friedmann cosmological spacelike initial
singularity in which the Ricci scalar diverges. 

Non-scalar polynomial curvature singularities (or `intermediate
singularities') are singularities where all the scalar
polynomials in the Riemann tensor are bounded, and on every curve ending at the
singularity (nonparallel) frames can be chosen in which the
components of the Riemann tensor are bounded.
Examples include the
cosmological `whimpers'  found in anisotropic SH
tilting fluid cosmologies \cite{EK74}. In addition, a
pp-wave spacetime can be singular, but all of its scalar
invariants are identically zero \cite{SKMHH}. These types of singularities
are more complicated since they need not be encountered by all
local observers.

It should be noted that the presence of a diverging scalar invariant
is not sufficient to prove the presence of a spacetime singularity
(and it may not be necessary). In the Kruskal-Szekeres spacetime,
which is the maximal analytic extension of the Schwarzschild metric,
the scalar polynomial formed from the first derivatives of the
curvature, $R^{abcd;e}R_{abcd;e}$, vanishes exactly at the horizon
\cite{KLA}. This can be used to illustrate that the existence of
invariants that blow up does not necessarily imply a spacetime
singularity (for example, consider $1/R^{abcd;e}R_{abcd;e}$)
\cite{MAC}.

If a spacetime is geodesically complete, it may still
be pathological or singular in some sense,
e.g., as in the examples above,
such a spacetime may still admit non-geodesic (i.e., accelerating)
timelike curves that are incomplete.
For this reason no definition distinguishing singular and non-singular
spacetimes by geodesic completeness alone can be totally satisfactory
\cite{Geroch}: in particular, worldlines that represent observers ought
to be complete, and geodesic completeness does not prevent other curves from
`escaping' the spacetime.
Classic examples include Geroch's rocket with
bounded acceleration, Misner's example of trapped curves in a
torus, and tilting perfect fluid observers (in the future
direction) in LRS Bianchi V models with unbounded acceleration. 
There are also examples in Minkowski spacetime in which
accelerating curves can be found so that the expansion blows up in
a finite proper time, and hence
no meaningful extension can be defined \cite{TCE}. 
It is of use to define singular spacetimes in such
a way that the pathological behaviour is clear and meaningful (and
carries information on how the singular behaviour occurs)
\cite{MAC}.
Spacetimes such that all curves ending at a
singularity have the property that the components of the Riemann
tensor in a parallelly propagated frame remain bounded along the
curve, have been called  `quasiregular' \cite{TCE}.

Therefore, a pathological spacetime
that does not admit a spacetime singularity must have a structure 
defined on the spacetime manifold
that is singular, and this may
or may not be accompanied by a divergence of the components of the
curvature tensor
or its derivatives. There could be (i) pathologies
of mathematical structures defined on the spacetime manifold
(e.g., poorly behaved congruences or spinor structure, conical
singularities), or  (ii) pathologies of physical structures defined
on the spacetime manifold (e.g.,  worldlines of matter). Singularities
for which physical quantities diverge
are perhaps of more 
importance. However, in the context of current theories of high-energy
physics,
lines may be blurred between geometrical and physical
singularities; e.g., in supergravity theory additional fields such as
the dilaton
and form fields are gravitational degrees of freedom.

In the previous section, we have shown that Bianchi V models provide
examples of such pathologies which are not spacetime singularities,
where components of high enough order covariant derivatives of the Riemann tensor blow up.
In this case the `singularities' in the derivatives are in
some sense at infinity, although reachable in finite time by the
fluid. The example discussed by Musgrave and Lake \cite{MusLak95}
shows that one can have similar blowup of derivatives, without a
spacetime singularity, at the centre of a spherically symmetric
configuration.

We now answer the question raised at the end of the introduction.
For the fluid congruence in the LRS Bianchi type V models, the kinematic singularity, which occurs
at the asymptotic state $M^-$ for $\frac43 < \gamma < 2$, is associated with the blow up of
components of covariant derivatives of the Weyl tensor of order
$n > \frac{3(2-\gamma)}{3\gamma-4}$,
and of the Ricci tensor of order
$n > \frac{2\gamma}{3\gamma-4}$.
That is, at the kinematic singularity, while the components of the Riemann tensor and
its low-order covariant derivatives are bounded,
those of its high-order covariant derivatives blow up.
We speculate that this is also true in general.

In the LRS Bianchi type V models,
the quantities which blow up are Cartan invariants defined by taking a
frame defined by the fluid motion (up to the spacetime isotropy). In
\cite{inv} it was shown that for spacetimes not in the Kundt class the
components of the curvature tensors are, at least in principle,
determined by the polynomial curvature invariants.  The components
correspond to roots of a set of characteristic polynomials;
consequently, assuming that the algebraic type does not change, the
components will blow up if and only if the polynomial curvature
invariants blow up.  Therefore, except in Kundt spacetimes, there
exists a Cartan invariant that blows up if and only if there exists a
polynomial curvature invariant that blows up. Hence our results show
that, in general, one or more scalar polynomial invariants will blow up.%
\footnote{To prove this rigorously it is necessary to assume that 
not only the algebraic types 
(of the curvature tensor and its covariant derivatives)
do not change in the appropriate limit,
but also that the differences between particular eigenvalues 
of certain projective operators do not approach zero in this limit.}

Finally, there is the issue of the possible resolution of
classical singularities within quantum theory. 
At Planck scale curvatures, the character of gravity may
change radically due to its underlying quantum nature.
It is expected that
singularities will be `smoothed out' or `resolved' in the
correct theory of quantum gravity. For example, it is
possible that there are  regular solutions of the Dirac equation
that can be extended through any classical singularity, and
since strings experience
the spacetime only through the sigma model, spacetimes
which are singular in general relativity can be completely
nonsingular in string theory \cite{Horo}.

\section*{Acknowledgment}
This work was supported, in part, by NSERC of Canada.

\end{document}